# Superconductivity and Single Crystal Growth of $Ni_{0.05}TaS_2$

X.D. Zhu[a], Y.P. Sun[a,b,*], S.B. Zhang[a], H.C. Lei[a], L.J. Li[a], X.B. Zhu[a], Z.R. Yang[a], W.H. Song[a], J.M. Dai[a]

[a] Key Laboratory of Materials Physics, Institute of Solid State Physics, Chinese Academy of Sciences, Hefei 230031, People's Republic of China

[b] High Magnetic Field Laboratory, Chinese Academy of Sciences, Hefei 230031, People's Republic of China

**Abstract**

Superconductivity was discovered in $Ni_{0.05}TaS_2$ single crystal. $Ni_{0.05}TaS_2$ single crystal was successfully grown via the NaCl/KCl flux method. The obtained lattice constant *c* of $Ni_{0.05}TaS_2$ is 1.1999 nm, which is significantly smaller than that of $2H\text{-}TaS_2$ (1.208 nm). Electrical resistivity and magnetization measurements reveal that the superconductivity transition temperature of $Ni_{0.05}TaS_2$ is enhanced from 0.8 K ($2H\text{-}TaS_2$) to 3.9 K. The charge-density-wave transition of the matrix compound $2H\text{-}TaS_2$ is suppressed in $Ni_{0.05}TaS_2$. The success of $Ni_{0.05}TaS_2$ single crystal growth via the NaCl/KCl flux demonstrates that NaCl/KCl flux method will be a feasible method for single crystal growth of the layered transition metal dichalcogenides.

* Corresponding author. Tel.: +86 551 559 2757; Fax: +86 551 559 1434.

*E-mail address:* ypsun@issp.ac.cn

## 1. Introduction

Recently, several iron based superconductors *Ln*OFe*Pn* (*Ln*=Rare earth, *Pn* =pnictide) [1, 2], $A_x$FeAs with ThCr$_2$Si$_2$ structure (A is A = Ca, Sr, or Ba with substitution of A by an alkali metal such as K or Cs.) [3-5] and α-FeSe$_x$ [6] were discovered, which aroused a worldwide surge in searching for new superconductors. Interestingly, all of these compounds, as well as the copper-based high temperature superconductors [7, 8], are layered structure materials. Layered transition metal dichalcogenides (TMDC), such as TiSe$_2$, NbSe$_2$ and TaS$_2$, are well-known layered materials [9]. The structure of layered TMDC can be regarded as stacking of covalent coupled X-T-X (T, represents the transition metal, X = S, Se, Te) sandwiches, and the coupling between sandwiches is of weak van de Waals type. Organic molecules, alkali metals and transition metals can be intercalated into the van de Waals gap between the sandwiches [10-16]. 2H-TaS$_2$ undergoes a charge-density-wave (CDW) transition at ~ 78 K and a superconducting transition at ~ 0.8 K [17]. The intercalation effects in TaS$_2$ including superconductivity have attracted extensive efforts during the past decades [10, 14, 16, 18, 19, 20]. As for 3d-transitional metal (3d-TM), superconductivity has been observed in Fe dilute alloys Fe$_{0.05}$TaS$_2$ [14] and copper intercalated Cu$_x$TaS$_2$ [21, 22]. Recently, superconductivity has also been discovered in Cu$_x$TiSe$_2$ [23, 24]. Whether superconductivity can be discovered in other 3d-TM$_x$TaS$_2$ becomes an intriguing question. The Ni intercalated compound Ni$_{1/3}$TaS$_2$ has been reported in some prior work [25, 26], and no superconductivity has been discovered in Ni$_x$TaS$_2$ compounds. Thus, experiments were carried out to search for superconductivity in Ni$_x$TaS$_2$ with x < 0.1.

Chemical-vapor-transport (CVT) is a traditional method for the single crystal growth of TMDC [27]. Flux method, which had rarely been applied in single crystal growth of TMDC, was introduced by M. Y. Tsay et al. in single crystal growth of RuX$_2$ using Te as flux [28]. In this paper, we report the single crystal growth of Ni$_{0.05}$TaS$_2$ using the NaCl/KCl flux method and its enhanced superconductivity.

## 2. Experimental procedure

First, Ni$_{0.05}$TaS$_2$ powder was synthesized by heating stoichiometric Ni (99.9%), Ta (99.9%) and S (99.5%) powders in an evacuated quartz tube at 900°C for 4 days. The obtained Ni$_{0.05}$TaS$_2$ powder with 5 times NaCl/KCl flux in weight were mixed and grounded thoroughly. Then, the mixed powders were sealed in an evacuated quartz tube. The tube was placed in a vertical furnace, heated to 850°C and maintained for 6 hours. Subsequently, the tube is cooled to 650°C in a rate of 2°C/hour and kept at 650°C for 2 days before the following furnace cooling. The grown single crystals were separated by resolving the NaCl/KCl flux using deionized water. The dark blue, mirror-like plates in a typical size of 1.5 × 1.5 × 0.2 mm$^3$ were obtained, and the photograph is shown in the inset of Fig. 1. The 2$H$-TaS$_2$ and Cu$_x$TaS$_2$ single crystals mentioned in this paper were grown via the CVT method using iodine as an agent [22].

The composition of Ni$_{0.05}$TaS$_2$ single crystal was checked by energy dispersive x-ray spectroscopy (EDS). Some single crystals were crushed into powder for the powder X-ray diffraction pattern (XRD) experiments. The single crystal XRD and powder XRD experiments were performed on a Philips X'pert PRO x-ray diffractometer with Cu K$_\alpha$ radiation at room temperature. Ni$_{0.05}$TaS$_2$. The electrical resistivity experiments were performed by the standard four-probe method using a quantum design physical property measurement system (PPMS) (1.8 K ≤ T ≤ 400 K, 0 T ≤ H ≤ 9 T). Magnetization measurement as a function of temperature was performed in a quantum design superconducting quantum interference device (SQUID) system (1.8 K ≤ T ≤ 400 K, 0 T ≤ H ≤ 5 T).

## 3. Results and discussions

The EDS pattern for the Ni$_{0.05}$TaS$_2$ single crystal is shown in Fig. 1. The determined mol ratio of Ta : S is about 1 : 2. The nickel peak is evident in the EDS spectrum, and the determined mol ratio of the Ni : Ta is about 0.05 : 1, which is consistent with the nominal ratio of the raw materials.

Fig. 2 shows the single crystal XRD patterns for Ni$_{0.05}$TaS$_2$ and 2H-TaS$_2$ single crystals. The 2$\theta$ peak positions of Ni$_{0.05}$TaS$_2$ and 2$H$-TaS$_2$ are very close. This suggests that they have the same crystal structure and that their crystal surfaces are both parallel to (*00l*) crystal planes. The inset of Fig. 2 shows the magnification plots of the single crystal XRD patterns for Ni$_{0.05}$TaS$_2$ and 2$H$-TaS$_2$. Obviously, compared with the 2$\theta$ peak positions of 2H-TaS$_2$, the 2$\theta$ peak positions of Ni$_{0.05}$TaS$_2$ shift to higher angle. The single XRD pattern for Ni$_{0.05}$TaS$_2$ reflects that the single crystal Ni$_{0.05}$TaS$_2$ is of good quality. Fig. 3 shows the powder XRD patterns for Ni$_{0.05}$TaS$_2$ and 2H-TaS$_2$. Apparently, the powder XRD patterns confirm that Ni$_{0.05}$TaS$_2$ and 2H-TaS$_2$ have the same structure. Meanwhile, the shift of the 2$\theta$ peak positions of Ni$_{0.05}$TaS$_2$ can be also observed in the inset of Fig. 3. According to the Bragg formula $2d\sin\theta = n\lambda$, the shift of the (*00l*) peak positions demonstrate the lattice constant *c* is reduced in $Ni_{0.05}TaS_2$. In contrast, the copper intercalation in $Cu_xTiSe_2$ [23] and $Cu_xTaS_2$ [21, 22] result in the expansion of the lattice constant *c* with increase of *x* when *x* < 0.1. The reduction of lattice constant *c* suggests that Ni ions occupy the positions of Ta in $Ni_{0.05}TaS_2$.

The structural parameters of $Ni_{0.05}TaS_2$ were refined by the standard Rietveld technique [29]. Although the exact position of Ni ions can not be determined due to the low concentration and random distribution, better refinement can be obtained if Ni ions are assumed to occupy the position of Ta. This supports that Ni ions occupy the positions of Ta in $Ni_{0.05}TaS_2$. Fig. 4 shows the calculated XRD and the experimental XRD patterns. The structure parameters obtained from the best refinement are listed in Tab. 1. Due to the prefered orientation of (*00l*) crystal planes during the XRD experiment, the best $R_p$ value is relatively larger. The determined lattice constant *c* of $Ni_{0.05}TaS_2$ is 1.1999(1) nm, while that of 2H-$TaS_2$ is 1.208 nm [22].

Fig. 5 shows the temperature dependence of electrical resistivity in *ab* plane ($\rho_{ab}$-*T*) for $Ni_{0.05}TaS_2$. The $\rho_{ab}$-*T* curve shows a metallic behavior in the normal state and a nearly linear relation can be observed in the temperature range of *T* > 100 K. The resistivity starts to drop at *T* = 3.9 K, which indicates the superconductivity with $T_{Conset}$ = 3.9 K. The inset shows the enlarged view of the $\rho_{ab}$-*T* curve near the superconducting region. The resistivity drops to zero at *T* = 2.2 K with a transition width (10-90%) of about 1.4 K. The wide superconducting transition has been also observed in $Fe_{0.05}TaS_2$ [14]. The large transition width may originate from the inhomogeniety.

Fig. 6 shows the temperature dependence of dc magnetic susceptibility for $Ni_{0.05}TaS_2$. The magnetic field of 10 Oe was applied parallel to the *c* axis of a $Ni_{0.05}TaS_2$ plate. A sharp drop of magnetization at 2.2 K was observed for zero-field-cooling (ZFC) measurements，which further confirms the existence of superconductivity. The magnetic susceptibility does not saturate in the experimental temperature range.

Fig. 7 depicts the normalized resistivity ($\rho_{ab}(T)/\rho_{ab}(300K)$) for $Ni_{0.05}TaS_2$, $Cu_{0.08}TaS_2$, $Cu_{0.03}TaS_2$ and 2H-$TaS_2$, respectively. The enlarged view of the low temperature region is shown in the inset of Fig. 7. Kinks can be observed in the resistivity curves for $Cu_{0.03}TaS$ and 2H-$TaS_2$, which reflects the CDW transitions [22]. While no sign of CDW transition can be observed for $Ni_{0.05}TaS_2$ and $Cu_{0.08}TaS_2$. The residual resistivity ratio (RRR), $\rho_{ab}(300K)/\rho_{ab}(5K)$ for $Ni_{0.05}TaS_2$ is 1.6, which is smaller than 3.1 for $Cu_{0.08}TaS_2$. The low RRR has been also observed in graphite intercalated compounds $CaC_6$ (RRR ≈ 2) [30] and $Cu_xTiSe_2$ [23]. The scattering between the conducting carriers and the foreign atoms may be the origins of the observed low RRR in various intercalated compounds.

Combined with superconductors $Cu_xTiSe_2$ [23], $Fe_{0.05}TaS_2$ [14] and $Cu_xTaS_2$ [21, 22], the superconductivity discovered in $Ni_{0.05}TaS_2$ implies that superconductivity may exist in other 3d-TM intercalated or dilute doped TMDC with low concentration. In addition, NaCl/KCl flux method can be a feasible method for single crystal growth of the layered TMDC.

## 4. Conclusion

In summary, $Ni_{0.05}TaS_2$ single crystals were successfully grown via the NaCl/KCl flux method. Structure analysis shows that the lattice constant *c* of $Ni_{0.05}TaS_2$ is smaller than that of $2H-TaS_2$，which may originate from the Ta replacement by Ni. Electrical resistivity and magnetization measurements reveal that the superconductivity transition temperature of $Ni_{0.05}TaS_2$ is enhanced from 0.8 K ($2H-TaS_2$) to 3.9 K. The CDW transition of the matrix compound $2H-TaS_2$ is suppressed in $Ni_{0.05}TaS_2$. The observation of superconductivity in $Ni_{0.05}TaS_2$ implies that superconductivity may exist in other 3d TM intercalated or dilute doped TMDC with low concentration. In addition, the NaCl/KCl flux method provides a potential method for single crystal growth of the layered transitional metal dichalcogenides.

**Acknowledgements**

This work is supported by the National Key Basic Research under contract No. 2006CB601005, 2007CB925002, and the National Nature Science Foundation of China under contract No.10774146, 10774147 and Director's Fund of Hefei Institutes of Physical Science, Chinese Academy of Sciences.

Table captions:

Table 1. The refined structure parameters of $Ni_{0.05}TaS_2$: Space group (S.G.), lattice constant $a$, lattice constant $c$, $R_p$ value, $R_{wp}$ value, and the atom positions.

Tables:

| Structure parameters of $Ni_{0.05}TaS_2$ | |
|---|---|
| S.G. | $P6_3/mmc$ |
| $a$ (nm) | 0.33128(2) |
| $c$ (nm) | 1.1999(1) |
| $R_p$ | 11.1555 |
| $R_{wp}$ | 14.3405 |
| Ta/Ni | (0, 0, 1/4) |
| S | (1/3, 2/3, 0.1206) |

Table 1 X. D. Zhu et al.

Figure Captions:

Fig. 1. The EDS pattern for Ni$_{0.05}$TaS$_2$. Inset: photograph of the grown Ni$_{0.05}$TaS$_2$ single crystals.

Fig. 2. The single crystal XRD patterns for Ni$_{0.05}$TaS$_2$ and 2$H$-TaS$_2$. The inset shows the magnification plot of (006) diffraction peak.

Fig. 3. The powder XRD patterns for Ni$_{0.05}$TaS$_2$ and 2$H$-TaS$_2$. The inset shows the magnification plot of (006) diffraction peak.

Fig. 4. The experimental and calculated XRD patterns for Ni$_{0.05}$TaS$_2$. Cross: the experimental pattern; upper solid line: the calculated XRD pattern; bar: the calculated peak positions; the lowest line: the difference of the experimental and calculated XRD patterns.

Fig. 5. The temperature dependence of the in-plane resistivity ($\rho_{ab}$-$T$) for Ni$_{0.05}$TaS$_2$. The inset shows the $\rho_{ab}$-$T$ curve near the superconducting region. The arrow shows the zero resistance transition temperature.

Fig. 6. The temperature dependence of ZFC susceptibility measured in an applied field of 10 Oe parallel to $c$ axis for Ni$_{0.05}$TaS$_2$. The arrow shows the transition temperature $T_C$.

Fig. 7. The temperature dependence of the normalized resistivity ($\rho_{ab}(T)/\rho_{ab}(300K)$) for Ni$_{0.05}$TaS$_2$, Cu$_{0.08}$TaS$_2$, Cu$_{0.03}$TaS$_2$ and 2$H$-TaS$_2$. The inset shows the enlarged view of the low temperature.